\documentclass[preprint,showpacs,12pt]{revtex4}
\usepackage{amsmath,amssymb,amsfonts}
\usepackage{graphicx}
\usepackage{hhline}
\usepackage{bm}
\usepackage{amsmath}
\usepackage{epsfig}

\begin{document}         
\begin{flushleft}
CERN-PH-TH-2014-253\\
\end{flushleft}
\title{Neutrino versus antineutrino cross sections and CP violation}

\author {M. Ericson}
\affiliation{Universit\'e de Lyon, Univ. Lyon 1,
 CNRS/IN2P3, IPN Lyon, F-69622 Villeurbanne Cedex, France}
\affiliation{Physics Department, Theory Unit, CERN, CH-1211 Geneva, Switzerland}

\author {M. Martini}
\affiliation{Department of Physics and Astronomy, Ghent University, Proeftuinstraat 86, B-9000 Gent, Belgium}

 \begin{abstract}
We discuss the nuclear interactions of neutrinos versus those of antineutrinos, 
a relevant comparison for CP violation experiments in the neutrino sector. 
We consider the MiniBooNE quasielastic-like double differential neutrinos and antineutrinos cross sections 
which are flux dependent and hence specific to the MiniBooNE set-up. 
We combine them introducing their sum ($\nu+\bar\nu$) and their difference ($\nu-\bar\nu$). 
We show that the last combination can bring a general information, which can be exploited in other experiments, 
on the nuclear matrix elements of the axial vector interference term.  
Our theoretical model reproduces well the two cross sections combinations. This confirms 
the need for a sizeable multinucleon component in particular in the interference term.
\end{abstract}
\maketitle 
\section{Introduction}

One of the challenging goals of neutrino experiments is the detection of CP violation in the neutrino sector. 
A convincing  evidence would be the detection of an asymmetry between the oscillation rates of muon neutrinos and antineutrinos into electron neutrinos since, in the absence of CP violation, these  rates are the same. 
For this test  one needs a muonic neutrino beam and an antineutrino one. 
The electrons (positrons) produced in a far detector by the charged current interaction of the electron neutrinos 
(antineutrinos)  
are the signature for the $\nu_\mu \to \nu_e$  ($\bar{\nu}_\mu \to \bar{\nu}_e$) oscillation process, once the direct $\nu_e$ 
($\bar{\nu}_e$) background is eliminated. 
Several obstacles can stand on the way of the detection of CP violation through the $\nu$, $\bar{\nu}$ asymmetry. 
One is that the interactions of neutrinos and antineutrinos with any nucleus are not identical but they differ by the sign of the axial vector interference term, creating an asymmetry unrelated to CP violation and which has to be fully mastered. This is not a trivial task due to the complexity of the nuclear dynamics. 
It reflects in the neutrino interactions and may obscure the message that one wants to extract on the oscillation mechanism. 
One example concerns the role of the multinucleon emission process
which in a Cherenkov detector is misidentified for a quasielastic one \cite{Marteau:1999jp,Martini:2009uj}. 
This misinterpretation produces an apparent increase of the neutrino quasielastic cross section, at the origin of the so called axial mass anomaly detected in the MiniBooNE experiments \cite{AguilarArevalo:2010zc} as now widely accepted \cite{Martini:2009uj,Martini:2010ex,Amaro:2010sd,Nieves:2011pp,Bodek:2011ps,Martini:2011wp,Nieves:2011yp,Amaro:2011aa,Lalakulich:2012ac,Nieves:2013fr,Martini:2013sha,Martini:2014dqa,Megias:2014qva}. 
The detection of CP violation which involves a comparison between neutrinos and antineutrinos events
needs an even more detailed understanding of the multinucleon processes since it concerns the difference between neutrino and antineutrino cross sections. 
This is not a priori obvious and it is the object of the present article.

\section{Analysis and Results}
In order to illustrate the respective roles of the multinucleon components in the $\nu$ 
and $\bar\nu$ cross sections, as we have introduced in a previous work \cite{Martini:2010ex}, 
we start by giving below the following simplified expression 
(we remind however that for the actual evaluation we use a more complete one)
for the double differential neutrino or antineutrino cross sections on a  nuclear target such as $^{12}$C 
\begin{eqnarray}
\label{eq1:cross_section}
\frac{d^2\sigma}{d \cos\theta d \omega} & = & \frac{G_F^2 \, 
\cos^2\theta_c}{\pi} k_lE_l \, \cos^2\frac{\theta}{2} \, 
\left[ \frac{({\bf{q}}^2-\omega^2)^2}{{\bf{q}}^4}\,G_E^2 \, R_\tau +   \frac{\omega^2}{{\bf{q}}^2} \, G_A^2 \, R_{\sigma\tau (L)}  \right.\nonumber \\ 
& + & \left.
 2 \left( \tan^2\frac{\theta}{2}+ \frac{{\bf{q}}^2-\omega^2}{2 {\bf{q}}^2} \right ) 
\left( G_M^2 \, \frac{\omega^2}{{\bf{q}}^2} + G_A^2 \right)  R_{\sigma\tau (T)}\right.\nonumber \\ 
& \pm& \left. 2 \,  \frac{E_\nu + E_l}{M_N} \, 
\tan^2\frac{\theta}{2} \, G_A \, G_M \, R_{\sigma\tau (T)} \right],
\end{eqnarray}
where the plus (minus) sign applies to neutrinos (antineutrinos). 
In this Eq. (\ref{eq1:cross_section}) $G_F$ is the weak coupling constant, $\theta_c$ the Cabibbo angle, $G_E$, $G_M$ and $G_A$ are the nucleon electric, magnetic and axial form factors, $E_\nu$ and $E_l$ the initial and final lepton energies, $k_l$ the modulus of the final lepton momentum, $\omega$ and ${\bf{q}}$ the energy and the momentum transferred to the nucleus and $\theta$ is the lepton scattering angle.
The cross section on the nuclear target is expressed here 
in terms of the nuclear responses $R$ to probes with various couplings to the nucleon, isovector (index $\tau$) or isovector with isospin and spin coupling (index $\sigma \tau$). 
For the last responses the isovector spin coupling can be spin transverse, 
$\sigma\tau (T)$, or spin longitudinal, $\sigma\tau (L)$. The responses are ${\bf{q}}$ and $\omega$ dependent. The last term of Eq. (\ref{eq1:cross_section}) which 
changes sign between $\nu$ and $\bar\nu$ is the axial vector interference term, 
the basic asymmetry which follows from the weak interaction theory. 
It is expressed here in terms of isospin spin-transverse nuclear response. 
Its evaluation for complex nuclei is not trivial due to the presence of many-body effects. 
This is why it is important to obtain an experimental information on this term, one object of the present work. 

Now, for our description of the multinucleon component of the responses, we followed the experimental indications provided by the electron scattering data. 
In the transverse, or magnetic, data the dip between the quasielastic and the Delta part of the response is filled, 
which we interpreted as an indication for the presence of two nucleon emission \cite{Alberico:1983zg}. In the charge response instead this component is absent. 
With these indications, for neutrinos we have introduced  the multinucleon component  only in the spin isospin response, the pure isovector one keeping its quasielastic character. This difference is at the origin of the statement in  our work of Ref. \cite{Martini:2010ex} that the multinucleon effect should be  relatively less pronounced
 for antineutrinos than for neutrinos, the reason being that the isovector response which is free from multinucleon effect plays a larger relative role for antineutrinos due to the negative sign of the interference term. This argument obviously holds only in the kinematical regions where the influence of the isovector component is significant.
 
We have previously tested our model on the MiniBooNE data for the differential cross sections \cite{AguilarArevalo:2010zc,AguilarArevalo:2013hm}, independently for neutrinos \cite{Martini:2011wp} and antineutrinos \cite{Martini:2013sha}. 
The good fit of the data provided by our model did not contradict our  statement on the respective roles of np-nh. 
However in these works,  the test was performed separately for the neutrinos and  the antineutrinos and we did not specifically address the detailed comparison between the two. As this comparison is essential for the CP violation detection we want to investigate in the present work  this question in more details.  For Cherenkov detectors, such as the MiniBooNE or the Super-Kamiokande ones, are we able to describe quantitatively the  difference between neutrinos and antineutrinos nuclear interactions?
The wealth of experimental data which has accumulated in the last years by the MiniBooNE experiment allows an experimental test for this comparison and we will explore the message which can be drawn from these data. 
The quantity best suited  for this exploration is the measured double differential quasielastic-like (\textit{i.e.} which incorporates also  multinucleon states) cross section $\frac{d^2\sigma}{d\cos\theta dE_{\mu}}$ with respect to the muon energy $E_\mu$ and the muon emission angle $\theta$, which is not affected by the reconstruction problem of the neutrino energy \cite{Martini:2012fa},\cite{Lalakulich:2012ac},\cite{Nieves:2012yz,Lalakulich:2012hs,Martini:2012uc}. However it is flux dependent, with 
\begin{equation}
\label{eq2:sigma _flux_folded}
\frac{d^2\sigma}{d\cos\theta d E_{\mu}} = \int \frac{d^2\sigma}{d \cos\theta d \omega} |_{\omega =E _{\nu}- E _{\mu}} \, \Phi(E _{\nu}) dE _{\nu}
\end{equation}
where $\omega$ is the energy transferred to the nucleus and $\Phi(E _{\nu})$ is the neutrino (or antineutrino)  normalized  flux. 
Neutrino experiments measure $\frac{d^2\sigma}{d\cos\theta dE_{\mu}}$ while nuclear physics evaluations calculate the quantity $\frac{d^2\sigma}{d\cos\theta d\omega}$, which is the basic ingredient for all analysis on neutrino data. The flux dependence in $\frac{d^2\sigma}{d\cos\theta dE_{\mu}}$ is a priori  an obstacle to extract a universal comparison between the two cross sections for neutrinos and antineutrinos, applicable to CP violation experiments. 
For each set of flux profiles the measured differential cross sections are different. 
In particular the $\nu$ and $\bar\nu$ asymmetry for $\frac{d^2\sigma}{d\cos\theta dE_{\mu}}$ has two sources, one arises from the basic weak interaction theory, as given in Eq. (\ref{eq1:cross_section}). The second one which arises from the flux asymmetry has no universal character and is specific for each experiment. Is there nevertheless something general and informative in the MiniBooNE data? It is the question that we will address in the following.

Let us consider the following  combinations: the sum, $sum$, and the difference, $dif$, of the double differential cross sections for neutrinos and antineutrinos with respect to the lepton emission angle, $\theta$, and to the energy $\omega$ transferred to the nucleus
\begin{equation}
 sum (\cos\theta,\omega) =\frac{d^2\sigma_{\nu}} {d\cos\theta d\omega} +  \frac{d^2\sigma_{\bar\nu}}{d\cos\theta d\omega}
\end{equation}
while for the difference $dif$ the sign plus is changed to a minus one. 
Notice from Eq. (\ref{eq1:cross_section}) that the difference,  $dif$, contains only one term, the axial vector interference one. 
It is this quantity which governs the difference in cross sections of ${\nu}$ and ${\bar\nu}$ but it is not accessible directly from neutrino data since only derivatives with respect to the emitted lepton energy are  measurable.
We also introduce $S$ and $D$, the corresponding flux integrated quantities, which are instead experimentally accessible  in the MiniBooNE data,
 \begin{equation} 
\label{eq4:S}
  S (\cos\theta,  E_{\mu})  = \frac{d^2\sigma_{\nu}}{d\cos\theta dE_{\mu}}+ \frac{d^2\sigma_{\bar\nu}}{d\cos\theta dE_{\mu}}  
\end{equation}
\begin{equation}
\label{eq5:D}
  D(\cos\theta, E_{\mu}) = \frac{d^2\sigma_{\nu}}{d\cos\theta dE_{\mu}}-\frac{d^2\sigma_{\bar\nu}} {d\cos\theta dE_{\mu}}  
\end{equation} 
 in which  quantities such as $ \frac{d^2\sigma_{\nu}}{d \cos\theta d E_{\mu}}$ have been defined previously in 
Eq. (\ref{eq2:sigma _flux_folded}). The MiniBooNE experimental distribution of $D(\cos\theta, E_{\mu})$ 
has been given by Grange and Katori  \cite{Grange:2014tya} in a tridimensional plot.

\begin{figure}
\begin{center}
  \includegraphics[width=12cm,height=8cm]{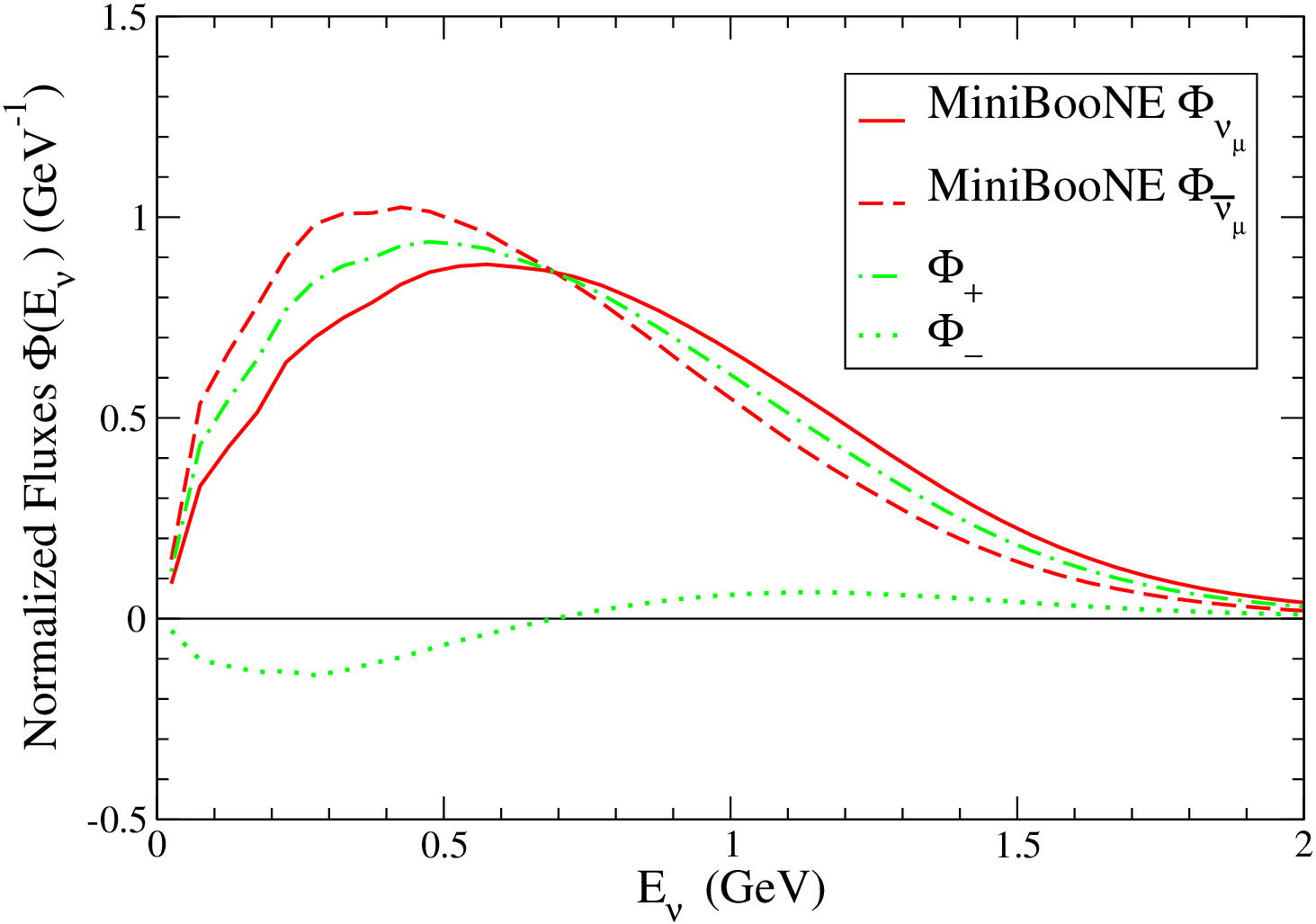}
\caption{(color online) Normalized MiniBooNE $\nu_\mu$ and $\bar{\nu}_\mu$ fluxes derived from Refs. \cite{AguilarArevalo:2010zc} 
and \cite{AguilarArevalo:2013hm} respectively. Their half sum ($\Phi_+$) and half difference ($\Phi_-$) are also shown.}
\label{fig_fluxes}
\end{center}
\end{figure}

If the normalized neutrino and antineutrino flux profiles would be identical, $\Phi_{\nu}(E_\nu) \equiv \Phi_{\bar{\nu}}(E_\nu)$, 
their common value could be factorized in the integrals over the neutrino energies  implicitly contained  in the above Eqs. (\ref{eq4:S}) and (\ref{eq5:D}). In this case  only the axial vector interference term would survive in the difference $D$, while the sum $S$ would  totally eliminate this term.
This would allow a direct experimental test of the interference part. However this is not quite the case as appears in Fig. \ref{fig_fluxes} 
which compares the two normalized MiniBooNE fluxes. In view of this difference we express the two cross section combinations in terms of the average flux, 
$\Phi_+ (E_\nu)= 1/2 [\Phi_{\nu}(E_\nu)+  \Phi_{\bar{\nu}}(E_\nu) ]$ and of the flux difference,  $\Phi_-(E_\nu) =  1/2 [\Phi_{\nu}(E_\nu)-  \Phi_{\bar{\nu}}(E_\nu)]$ 
 as follows
 \begin{equation}
\label{eq6:Sphi}
    S (\cos\theta,  E_{\mu}) = \int d E_\nu   \,[ sum(\cos\theta,\omega) |_{\omega =E _{\nu}- E _{\mu}} \Phi_+ (E_\nu)  +  
  dif(\cos\theta,\omega)|_{\omega =E _{\nu}- E _{\mu}} \Phi_- (E_\nu)]
\end{equation}
  and
\begin{equation} 
 \label{eq7:Dphi}
    D(\cos\theta, E_{\mu})=\int d E_\nu   \,[sum(\cos\theta,\omega)|_{\omega =E _{\nu}- E _{\mu}} \Phi_-  (E_\nu ) +  dif(\cos\theta,\omega)|_{\omega =E _{\nu}- E _{\mu}} \Phi_+(E_\nu)].
 \end{equation}
We observe again that for identical fluxes, for which $ \Phi_-  (E_\nu ) = 0$,   the quantity $D$ probes only the quantity $dif$ , i.e., the axial vector interference term, while the 
sum $S$ is totally blind to it. Notice that the MiniBooNE flux difference  $ \Phi_-  (E_\nu ) $ is negative for small $E_\nu $  values and becomes positive beyond an energy $\simeq .7 GeV$.

\begin{figure}
\begin{center}
  \includegraphics[width=16cm,height=12cm]{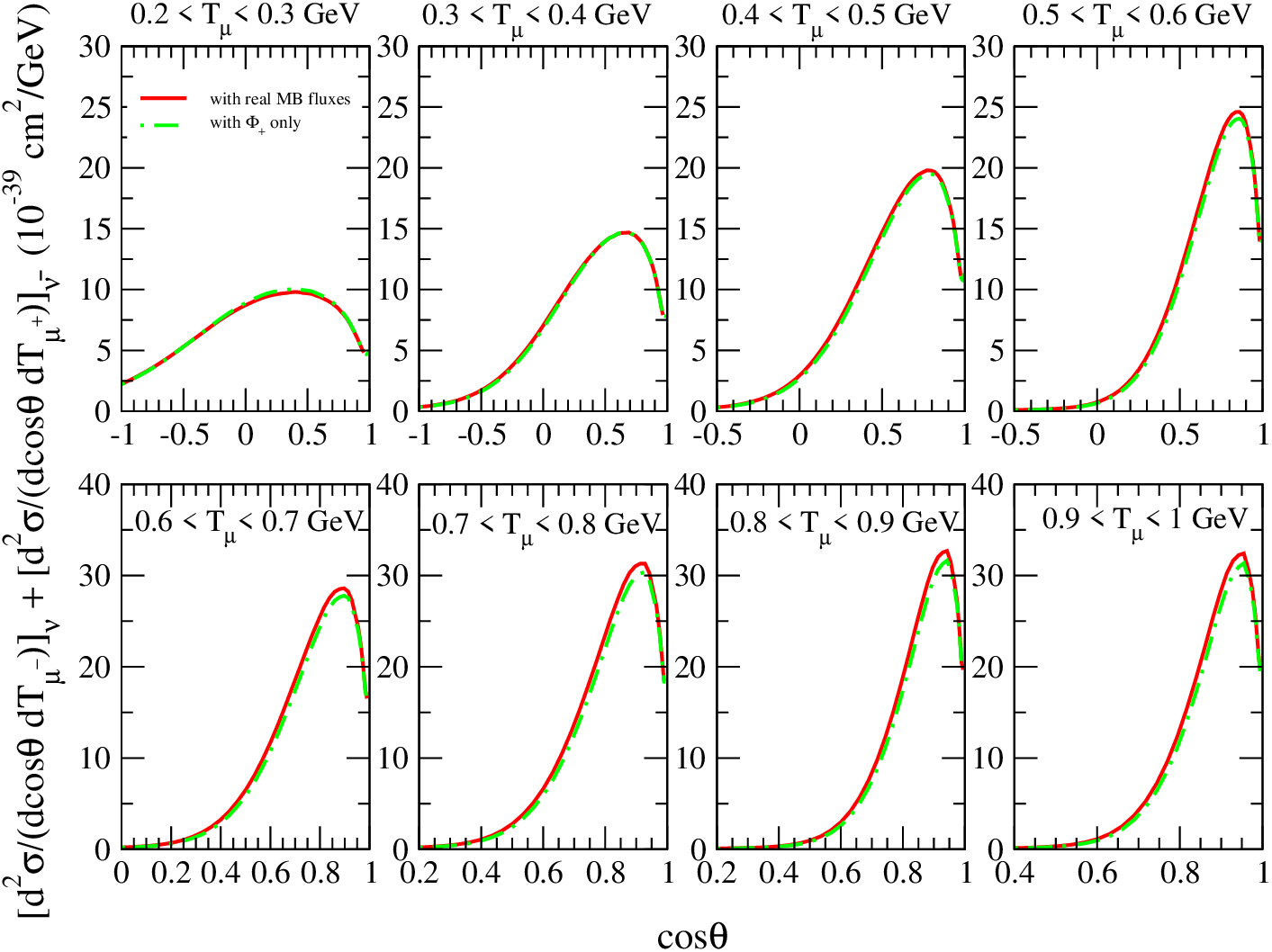}
\caption{(color online) Sum of the $\nu$ and $\bar{\nu}$ flux folded double differential cross sections on carbon per active nucleon plotted as a function of the $\cos\theta$ for different values of the emitted muon kinetic energies. 
Continuous line: evaluation (RPA + np-nh) with the real $\nu$ and $\bar{\nu}$ MiniBooNE fluxes; dot-dashed line: evaluation (RPA + np-nh) with the mean flux $\Phi_+$.}

\label{fig_sum_S_mean_flux}
\end{center}
\end{figure}

\begin{figure}
\begin{center}
  \includegraphics[width=16cm,height=12cm]{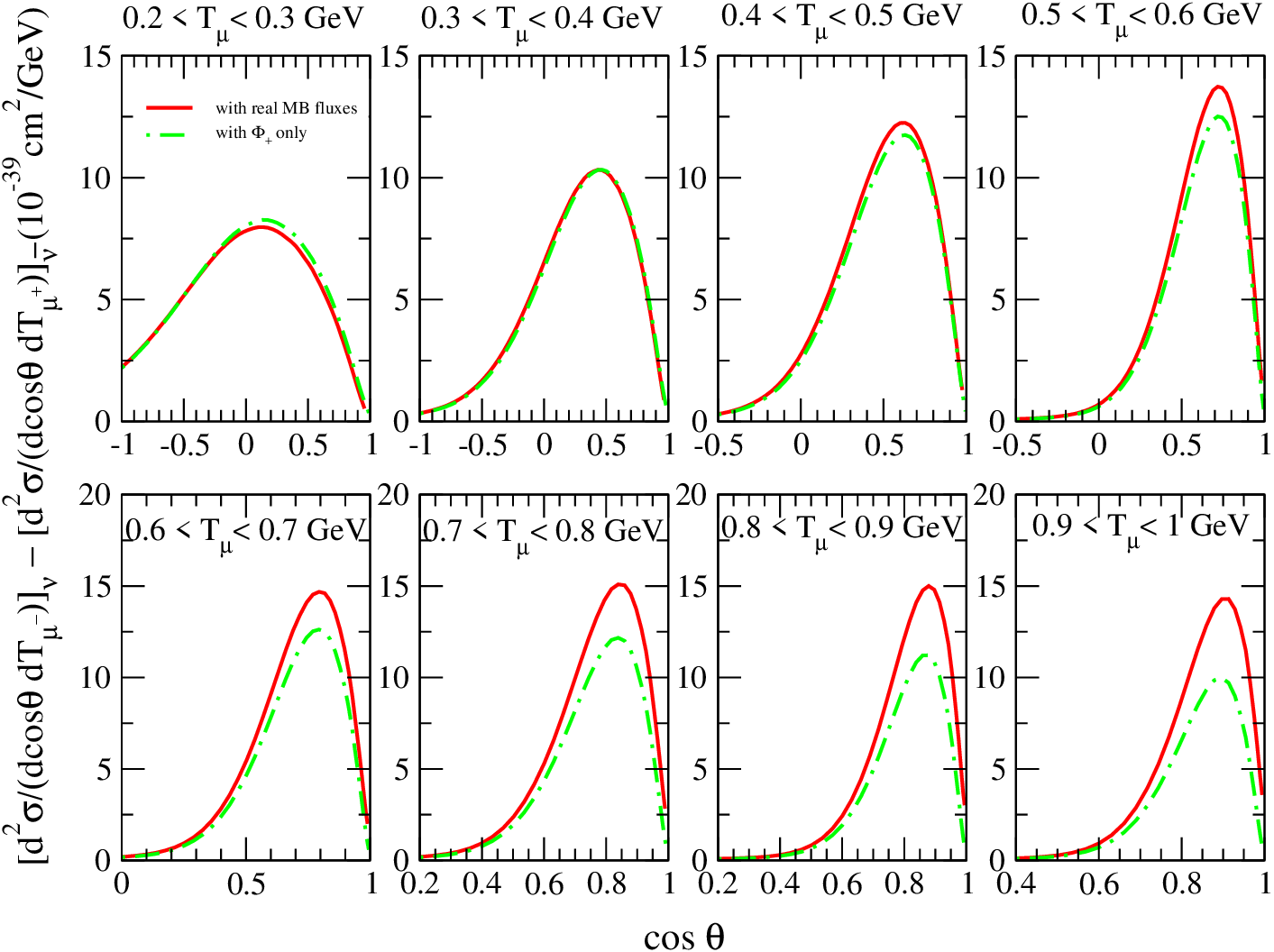}
\caption{(color online) Difference of the $\nu$ and $\bar{\nu}$ flux folded double differential cross sections on carbon per active nucleon plotted as a function of the $\cos\theta$ for different values of the emitted muon kinetic energies. 
Continuous line: evaluation (RPA + np-nh) with the real $\nu$ and $\bar{\nu}$ MiniBooNE fluxes; dot-dashed line: evaluation (RPA + np-nh) with the mean flux $\Phi_+$.}
\label{fig_dif_D_mean_flux}
\end{center}
\end{figure}
 
 The question is now the following: in view of the  fluxes difference, 
what remains in the MiniBooNE data of the purity of the difference $D$ with respect to the axial vector interference term and accordingly of its elimination in the sum $S$? 
In order to  answer these questions we have evaluated the sum $S$ and the difference $D$ with the real MiniBooNE fluxes on the one hand and  with only the mean flux $\Phi_+ (E_{\nu} )$ on the other hand. 
These evaluations are performed with our theoretical model which has been described for instance 
in Refs. \cite{Martini:2009uj,Martini:2011wp}. 
It incorporates the multinucleon component of the cross section and the  responses are treated in the random phase approximation, RPA. The resulting curves  are displayed in Figs. \ref{fig_sum_S_mean_flux} and  \ref{fig_dif_D_mean_flux} as functions of the muon angles, for various values of the muon kinetic energy. For what concerns the sum $S$, for all angles and in the full range of muon energies, the two sets of curves are  very close, confirming the major role of the mean flux contribution and showing that there is very little sensitivity in the sum on the axial vector interference term. 
For the difference $D$ the two sets of curves are quite close up to a muon kinetic energy of $  T_{\mu} \simeq .6$ to $.7$ GeV. 
Beyond this energy they progressively depart. 
The fact that the mean flux curve is below the genuine flux one reflects the positive sign of  $ \Phi_-  (E_\nu ) $ for large  $E_\nu $ values 
(large $T_{\mu}$ values meaning also large $E_\nu $ values).  
Notice also that for the smallest  $T_{\mu}$ values, for which the neutrino energy can be small, one can detect a change in the sign of the effect, which becomes slightly negative, reflecting the sign change of $ \Phi_-  (E_\nu )$. In spite of the (moderate) deviations at large muon kinetic energies, for all $T_\mu$ values the contribution from the mean flux remains dominant in the difference $D$. 
This is fortunate since it means that the axial vector interference term is experimentally accessible in the MiniBooNE data through the neutrino and antineutrino 
double differential cross sections difference. We can in particular test if our multinucleon component, which in our model is maximum in the axial vector interference term, i.e., in the difference $D$, is compatible, or not, with the data. 
While with the sum of the neutrino antineutrino cross sections instead we can explore its role in the remaining part of the cross section.
In general when the $\nu$ and $\bar\nu$ normalized flux difference is small the same conclusion applies: the difference in the measurable differential cross sections between neutrinos and antineutrinos is dominated by the axial-vector interference term. This is the case for T2K \cite{Abe:2012av} 
and also for the NuMI \cite{Anderson:1998zza} beams, the ones used in the MINOS, MINER$\nu$A and NO$\nu$A experiments.

\begin{figure}
\begin{center}
  \includegraphics[width=16cm,height=12cm]{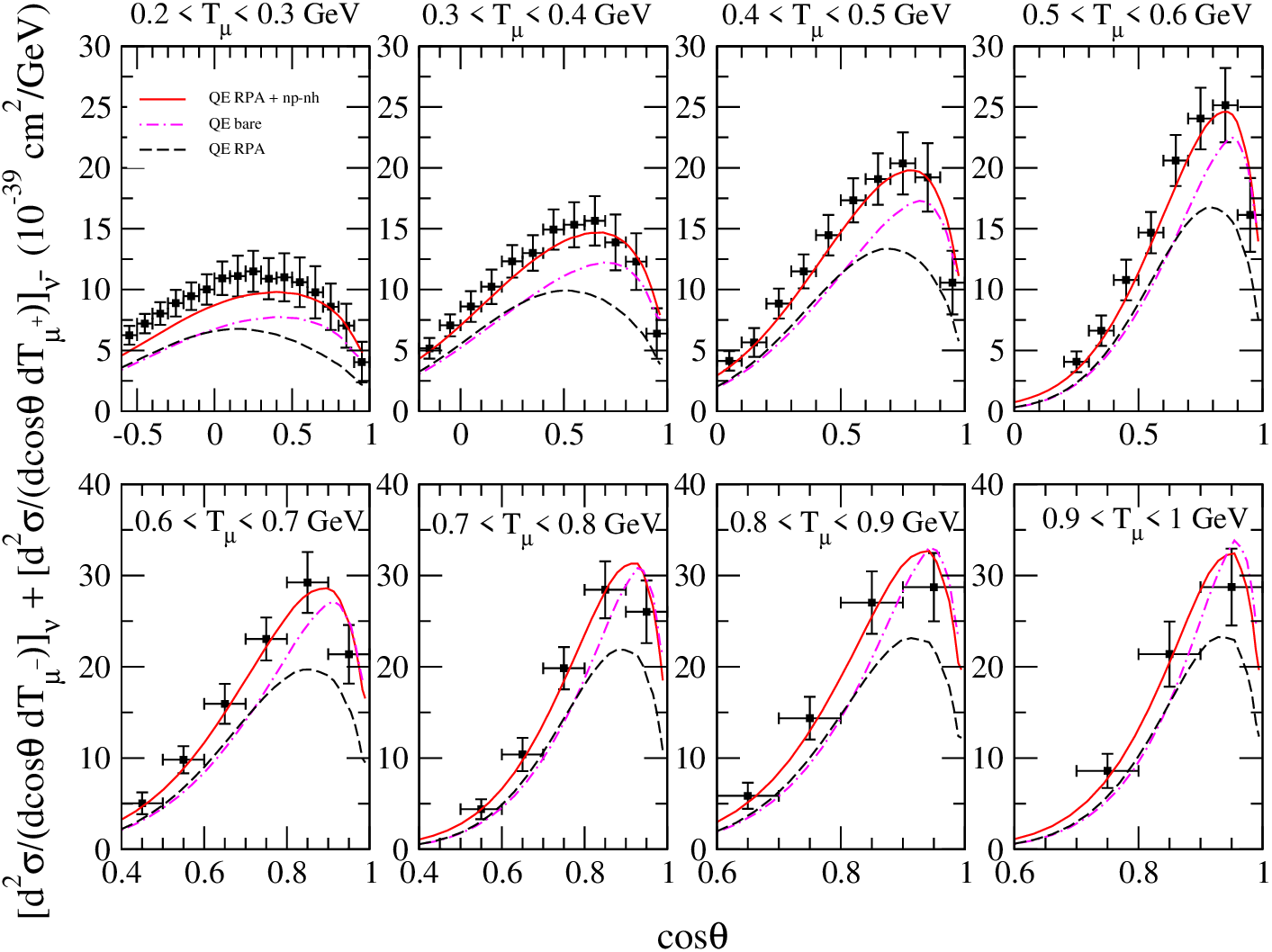}
\caption{(color online) Sum $S$ of the $\nu$ and $\bar{\nu}$ MiniBooNE flux folded double differential cross sections on carbon per active nucleon plotted as a function of the $\cos\theta$ for different values of the emitted muon kinetic energies. 
Continuous line: our complete RPA evaluation including the multinucleon emission channel; dashed line: genuine quasielastic contribution calculated in RPA; dot-dashed line: quasielastic contribution in the bare case. The points are the combination of the MiniBooNE experimental results \cite{AguilarArevalo:2010zc,AguilarArevalo:2013hm}. }

\label{fig_sum_S}
\end{center}
\end{figure}

\begin{figure}
\begin{center}
  \includegraphics[width=16cm,height=12cm]{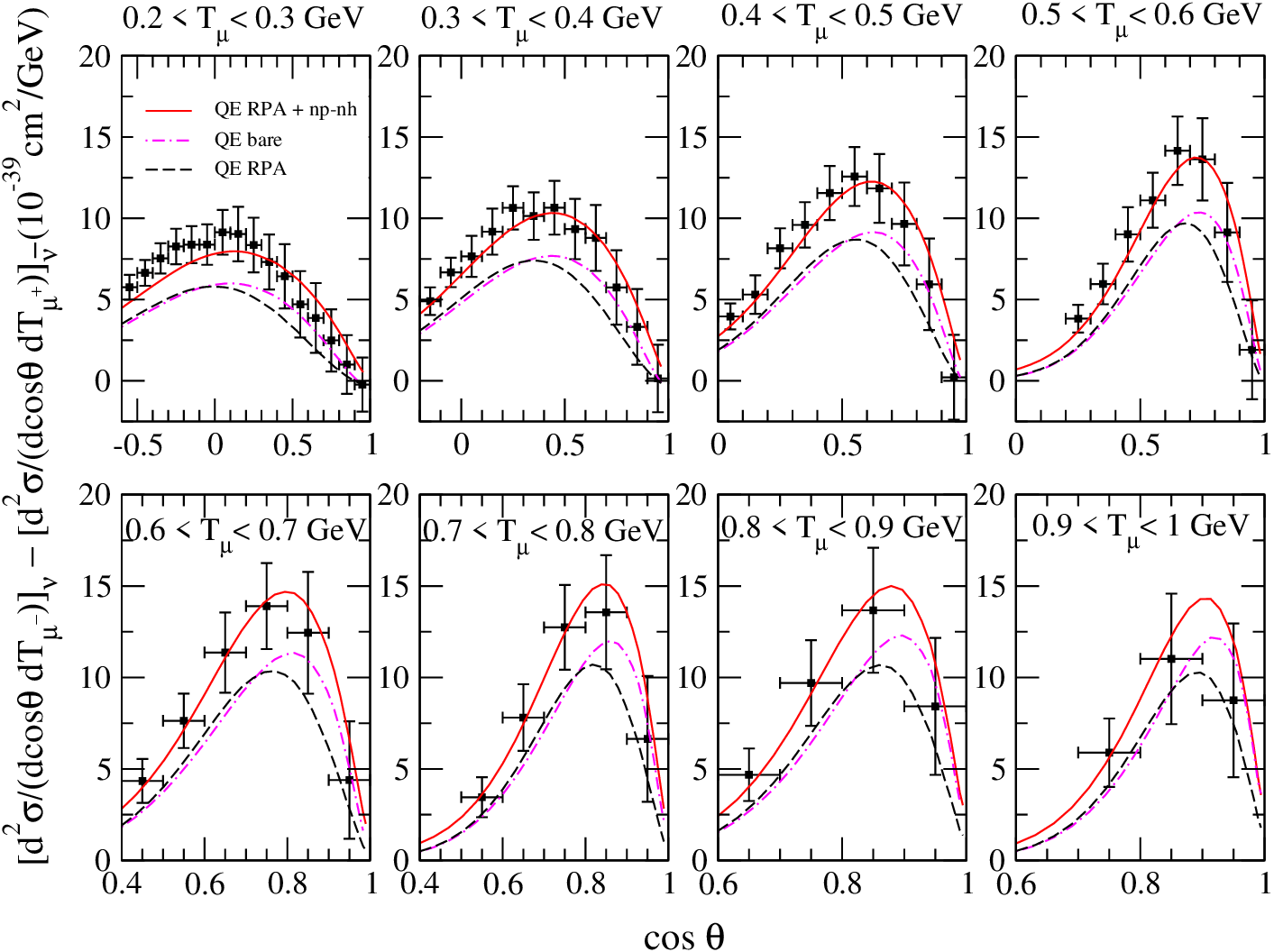}
\caption{(color online) The same of Fig. \ref{fig_sum_S} but for the difference $D$.}
\label{fig_dif_D}
\end{center}
\end{figure}

\begin{figure}
\begin{center}
  \includegraphics[width=15cm,height=10cm]{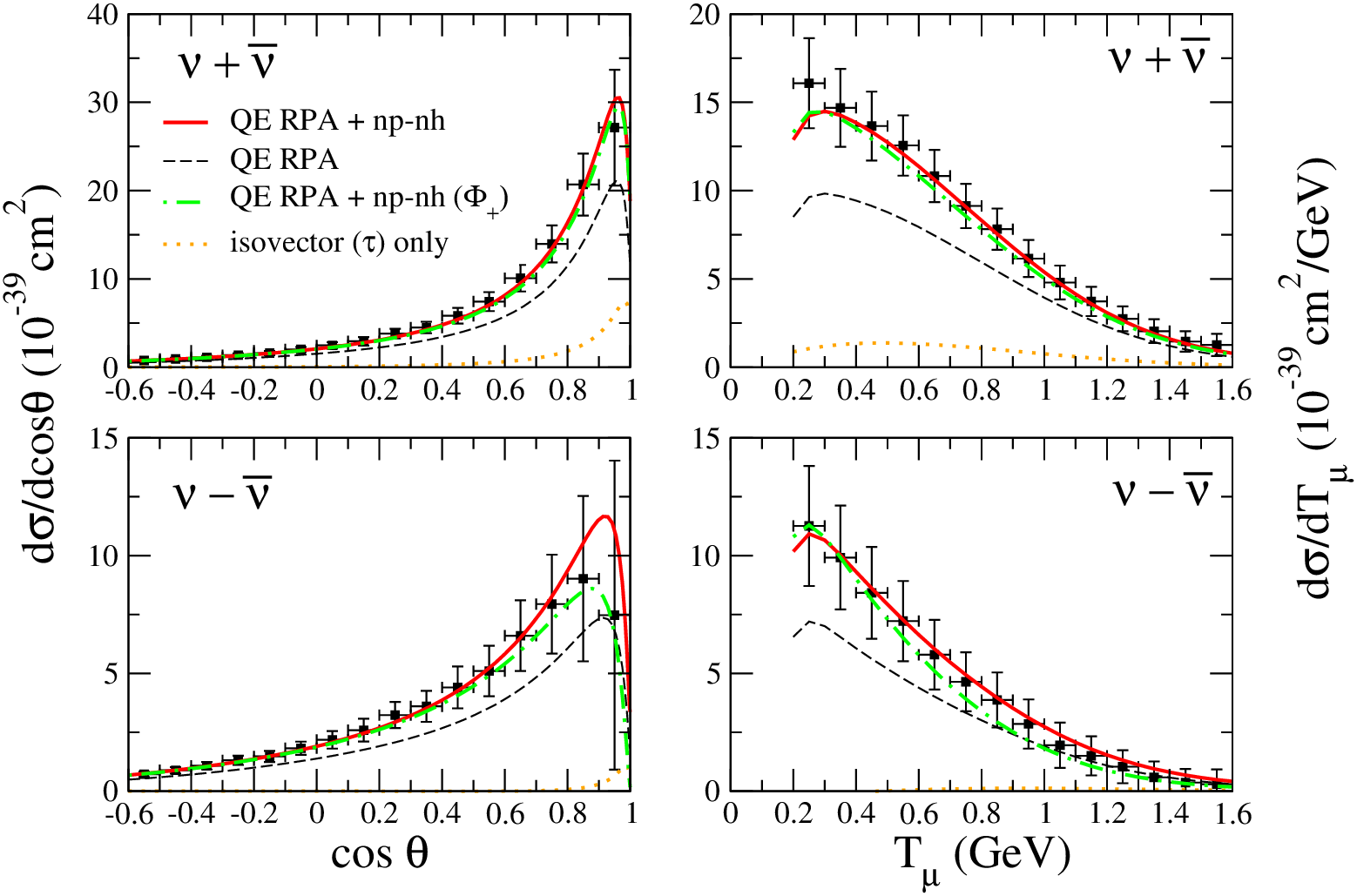}
\caption{(color online) Sum and difference of the $\nu$ and $\bar\nu$ flux folded differential cross sections $\frac{d\sigma}{d\cos\theta}$ 
and $\frac{d\sigma}{dT_\mu}$ on carbon per active nucleon. Continuous line: complete RPA evaluation including the multinucleon emission channel with the real $\nu$ and $\bar{\nu}$ MiniBooNE fluxes; dashed line: genuine quasielastic contribution calculated in RPA with the real $\nu$ and $\bar{\nu}$ MiniBooNE fluxes; dot-dashed line: complete RPA evaluation including the multinucleon emission channel with the mean flux $\Phi_+$; dotted line: contribution of the isovector response only.}
\label{fig_ds_cos_Tmu__SD}
\end{center}
\end{figure}

Having seen the message that the comparison with the mean flux curves carries \textit{i.e.} that the axial vector interference term dominates the difference and  has very little influence on the sum, in the following we calculate in our model the sum and the difference of the cross sections with the true neutrinos and 
antineutrinos fluxes, and not with the mean one, in order to avoid unnecessary errors.  Our present results are then simply the sum and the difference of our previous theoretical results published in Ref. \cite{Martini:2011wp} for neutrinos and in Ref. \cite{Martini:2013sha} for antineutrinos.
In Figs. \ref{fig_sum_S} and \ref{fig_dif_D} we display our calculated values of $S(\cos\theta,E_\mu)$ and $D(\cos\theta,E_\mu)$ as a function of the muon emission angle, for various values of the muon kinetic energy, together with the  experimental data points. Our predictions which incorporate the multinucleon component account quite well for the data for all angles and in the full range of muon energies, both for the sum and for the difference.  Only for the smallest $T_\mu$ bin some small deviations can be observed. 
   
This test of the validity of our model for these combinations is important as it addresses directly to the question of our understanding of the neutrinos versus antineutrinos interactions, in particular for what concerns the crucial and more debated role of the multinucleon component. 
Our predicted cross sections without the multinucleon part  are also shown in  Figs. \ref{fig_sum_S} and \ref{fig_dif_D}, they definitely fail to account for the data both for the sum and the difference. 
The necessity of the introduction of a large multinucleon component, in particular in the difference \textit{i.e.} 
in the axial vector interference term, is confirmed here by the curves of Fig. \ref{fig_dif_D}. 
Another check remains to be performed to fully confirm the necessity of the multinucleon piece, in particular in the axial vector interference term: 
in our description the collectivity of the quasielastic responses has been  included in the form of the random phase approximation treatment, RPA. 
It produces a suppression effect of the quasielastic part due to the repulsive character of the particle-hole force \cite{Alberico:1981sz}. For the spin-isospin transverse response it is the Ericson-Ericson--Lorentz-Lorenz quenching arising from the mixture of Delta-hole states into nucleon-hole ones \cite{Ericson:1966fm}. 
The question is then: if a good fit is obtained with the combined and opposite effects of RPA and of the multinucleon  component, could it be that a similar good fit would be achieved by  omitting both effects. Would the simplest quasielastic description also account for the data? In Figs. \ref{fig_sum_S} and \ref{fig_dif_D} the effect of RPA is suppressed in the quasielastic cross section, which indeed has some enhancement effect. In particular for the cross section difference this enhancement is moderate and not enough to account by itself, for the data. We can safely conclude that a large multinucleon component is needed to describe the data for the axial vector interference term which governs the cross sections difference. The same conclusion applies to the cross sections sum, i.e. to the remaining part of the interaction, it is also appreciably influenced by the multinucleon component.  
Our model for the neutrino nucleus interaction is able to describe both components. 
As an additional illustration of its success we report in Fig. \ref{fig_ds_cos_Tmu__SD} the single differential cross sections, with respect to the muon kinetic energy or to the muon emission angle. As previously we deal with the sum and the difference for neutrinos and antineutrinos calculated with the true and the averages fluxes. We can observe again that the sum shows practically no sensitivity to the flux  difference, while the cross sections difference displays some  a mild sensitivity, in particular in the forward direction or at large $T_\mu$ values. Our predictions which include the multinucleon component reproduce well the data. 
 In our model the relative importance of the multinucleon term in the cross sections combinations depends on the role of the isovector response 
which is shown in Fig. \ref{fig_ds_cos_Tmu__SD}, the smaller this role the larger the multinucleon contribution.  It is the largest in the cross section difference. Similarly the isovector response weight is larger  in antineutrino cross sections than in neutrino ones, hence the smaller multinucleon contribution for antineutrinos. It is  
 however  not a large difference  and  the multinucleon influence remains important also for antineutrinos. 

\begin{figure}
\begin{center}
  \includegraphics[width=16cm,height=6cm]{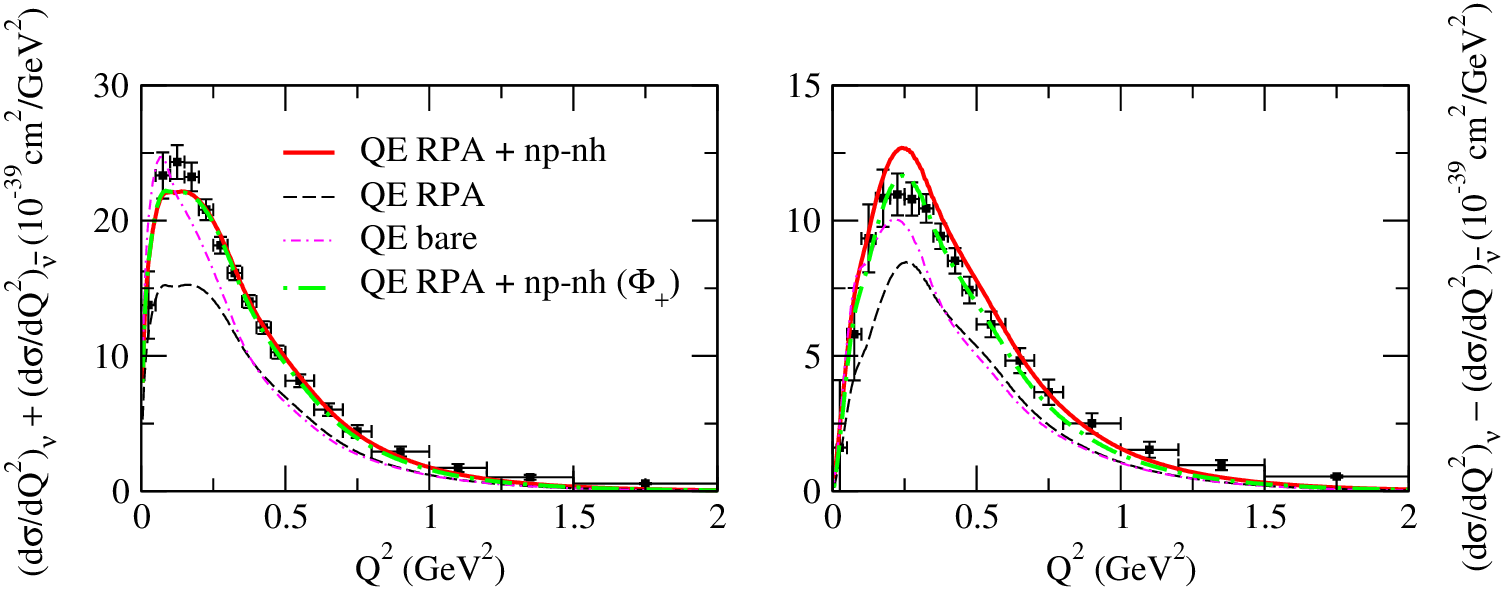}
\caption{(color online) Sum and difference of the $\nu$ and $\bar\nu$ flux folded $Q^2$ distributions on carbon per active nucleon. Continuous line: complete RPA evaluation including the multinucleon emission channel with the real $\nu$ and $\bar{\nu}$ MiniBooNE fluxes; dashed line: genuine quasielastic contribution calculated in RPA with the real $\nu$ and $\bar{\nu}$ MiniBooNE fluxes; thin dot-dashed line: quasielastic contribution calculated in the bare case with the real $\nu$ and $\bar{\nu}$ MiniBooNE fluxes; thick dot-dashed line: complete RPA evaluation including the multinucleon emission channel with the mean flux $\Phi_+$.}
\label{fig_Q2_SD}
\end{center}
\end{figure}

Finally for completeness we combine in Fig. \ref{fig_Q2_SD} our previous evaluations of the neutrino and antineutrino $Q^2$ distributions, published in Refs. \cite{Martini:2011wp} and \cite{Martini:2013sha} respectively, to evaluate their sum and difference. We also display the result obtained with the averaged flux $\Phi_+$. As previously pointed out the difference is more sensitive to the mean flux approximation which however gives the bulk of this difference. As a consequence the difference of the experimental MiniBooNE points $\left(\frac{d\sigma}{dQ^2}\right)_\nu-\left(\frac{d\sigma}{dQ^2}\right)_{\bar{\nu}}$ is directly related to the $Q^2$ distribution of the axial vector interference term. This conclusion would apply as well to the MINER$\nu$A neutrino \cite{Fiorentini:2013ezn} and antineutrino \cite{Fields:2013zhk} 
$Q^2$ distributions, due to the closeness of the neutrino and antineutrino normalized fluxes. It will be the object of a future investigation.

Notice that the present study is done for the muonic neutrinos of the MiniBooNE experiment, for which data are available while CP violating experiments through the asymmetry of the oscillations rates of muonic neutrinos and antineutrinos into electron ones involve the detection of electrons or positrons produced in a detector by the charged current interactions of these electron neutrinos. We have already addressed the question of electron neutrino cross sections, versus the muon neutrino ones \cite{Martini:2012uc}. The effect of the small change in kinematics due to the smaller electron mass  will not affect the present conclusions: 
the electron neutrino nuclear  interactions produce by themselves an important $\nu_e$ $\bar\nu_e$ asymmetry. The present study indicates that the multinucleon role is essential in this problem and to what precision this asymmetry can be mastered.

\section{Summary and conclusions}
 In summary we have investigated two different combinations of the neutrino and antineutrino MiniBooNE flux folded double differential  cross sections on $^{12}$C, 
their sum and their difference. They probe different pieces of the neutrino or antineutrino interactions with the nucleus. 
These quantities depend on the neutrino or antineutrino flux. In the case of identical  neutrino and antineutrino normalized flux profiles, the difference provides a direct access to the axial vector interference term, while the sum totally eliminates it. As this is not actually the case for the MiniBooNE fluxes we have tested how much the flux difference influences the two combinations. 
We have shown that this  influence is  small both on the sum and on the difference of the cross sections. These combinations remain rather pure with respect to the axial vector interference term which is either dominant (difference) or nearly absent (sum). This allows more specific tests of our theoretical model on the axial vector interference term, important for the CP violation data.
 Our model gives a good fit for the MiniBooNE data for the sum and the difference of the cross sections  reproducing well the data in the full range of muon energy and emission angle. The introduction of the multinucleon component is necessary for a good fit, an important test for its presence in the axial vector interference term.  The success of our description indicates that we can reach a good understanding of the nuclear effects in neutrino interactions, also for what concerns the comparison between neutrinos and antineutrinos cross sections. We have concentrated in this work on the interactions of muonic neutrinos where a complete set of data is available. Our predictions can easily be extended to electron neutrinos, relevant for CP violation data. 
The nuclear cross section difference for neutrinos and antineutrinos stands as a potential obstacle in the interpretation of  experiments aimed at the measurement of the CP violation angle, $\delta$.  The present analysis, performed on the MiniBooNE data, shows the importance, for Cherenkov detectors, of the inclusion of the multinucleon contribution to the quasielastic-like cross section for mastering this difference.

\begin{acknowledgements}
 This work was partially supported by the Interuniversity Attraction Poles Programme initiated by the Belgian Science Policy Office (BriX network P7/12).
\end{acknowledgements}

\end{document}